\documentclass[preprint,preprintnumbers,amsmath,amssymb,floatfix]{revtex4}
\usepackage{graphicx}
\usepackage{dcolumn}
\usepackage{bm}
\usepackage{epsfig}

\begin{document}

\centerline{\normalsize DESY 20-180 \hfill ISSN 0418--9833}

\title{Production of massless charm jets in $pp$ collisions at 
next-to-leading order of QCD in comparison with CMS data}



\author{Gustav Kramer}

\email{gustav.kramer@desy.de}

\affiliation{{II.} Institut f\"ur Theoretische Physik,
Universit\"at Hamburg, Luruper Chaussee 149, 22761 Hamburg, Germany}

\date{\today}

\begin{abstract}

We present predictions for the inclusive production of charm jets in 
proton-proton collisions at 2.76 and 5.02 TeV. The charm-quark is considered 
massless.
In this scheme we find that the ratio of the next-to-leading order to the 
leading order cross section (K factor) is almost equal to one depending 
essentially on the choice of the renormalization and factorization
scale. Adding non-pertubative corecctions obtained from Pythia Monte Carlo
calculations lead to resonable agreement with experimental c-jet cross sections
obtained by the CMS \cite{10} collaboration.

\end{abstract}

\maketitle
\thispagestyle{empty}

\section{Introduction}

The cross-section for producing jets in pp collisions at the Large Hadron
Collider (LHC) yields an important testing field of perturbative Quantum
Chromodynamics (QCD) and offers the possibility to learn more about
proton-parton distributions (PDFs). So far most of the predictions for LHC
experiments have been done for the sum of ingoing mass less quarks up to b
quarks. They have been found in very good agreement with the measurements. In
these predictions, the contributions of heavy quarks (charm and bottom)
constitute only a few percent of the total production cross-sections.
Considering the experimental data, from these comparisons it is not clear 
whether the heavy quark jets are satisfactorily described by perturbative QCD 
(pQCD) calculations.

So far, most of the predictions for heavy-quark jets separately have been done
in the massive quark scheme or fixed-flavor-number scheme (FFNS) \cite{1, 2} in
which heavy quarks appear in the final-state only and not as partons in the
initial state, as for example in the zero-mass variable-flavor-number scheme
(ZM-VFNS). In this scheme the calculations for the sum of all flavors have been
done. This is also the appropriate scheme for most of the PDF parametrizations 
of the proton. A few years ago we have presented predictions of massless charm
\cite{3} and bottom jets \cite{4} in $p\bar{p}$ and $pp$ collisions at 
next-to-leading order (NLO). In particular, in the latter reference, we have 
compared our predictions for bottom jets in $p\bar{p}$ collisions at $\sqrt{S} 
= 1.96$ TeV and $pp$ collisions at $\sqrt{S} = 7$ TeV. The measurements, we 
have been compared to, have been done by the CDF collaboration \cite{5, 6, 7} 
at the Tevatron and by the CMS collaboration \cite{8} at the LHC. At the LHC 
also the ATLAS collaboration has measured inclusive b-jet cross section at 
$\sqrt{S}= 7$ TeV \cite {9}, to which we have not compared our results, however.

In our work \cite {4} we have found that at small transverse momentum $p_T$ of 
the bottom jets the ratio of the NLO to the leading-order (LO) cross section is
smaller than one. It increases with increasing $p_T$ and approaches one at 
larger $p_T$ at a value depending essentially on the choice of the 
renormalization scale. Adding non-perturbative corrections obtained from 
PYTHIA Monte Carlo computations which contain in addition to the LO 
perturbative contributions the hadronic and parton shower corrections, we 
obtained reasonable agreement with the experimental b-jet across sections as 
measured by the CMS collaboration \cite{8}.

At the time of our earlier work \cite{4} data about charm-jet production,
neither from CDF nor by one of the LHC experiments were not available, 
presumably since the experimental identification of charm jets is much more 
difficult than for bottom-jets. Therefore we have not checked whether our 
findings for bottom-jet in \cite{4} is also true for the production of 
charm-jets. This is the purpose of this work, since very recently, experimental
data about single-inclusive charm-jet cross section have become available as 
presented by the CMS collaboration \cite{10} at the LHC. These cross-section 
have been measured for two $\sqrt{S}$ values, $\sqrt{S} = 2.76$ TeV and 
$\sqrt{S}= 5.02$ TeV for the rapidity region $|y| < 2.0$ and for five $p_T$ 
bins in the range $40 < p_T < 250$ GeV (in the case of the $\sqrt{S} =2.76$ TeV) and for $80 < p_T < 400$ GeV (for the case $\sqrt{S} = 5.02$ TeV). We have 
calculated the cross-section for the same $p_T$  bins as in the CMS experiment 
and the same rapidity ($y$) bin for both center-of-mass energies.

In Sec. 2 we describe the PDF input and outline the theoretical framework.
Section 3 contains our results for single-inclusive charm-jet cross-sections
$d\sigma/dp_T$ and the comparison with the cross-section measured by the CMC
collaboration. A summary and some conclusions are presented in Sect. 4.

\section{PDF Input and Theoretical Framework}

As in our earlier publications on charm jets \cite{3} and bottom jets \cite{4}
we rely on previous work on dijet production in the process $\gamma + p
\to jet + X$ \cite{11, 12} in which cross-sections for inclusive one-jet and 
two-jet production up to NLO for both, the direct and the resolved 
contributions, are calculated. The resolved part of this routine can be used 
for $pp$ collisions replacing the photon PDF by the proton PDF. The routine 
\cite{11, 12} contains quarks of all flavors up to and including the bottom 
quark as well as the gluon. In this work about charm jets the bottom quark will
be excluded. The routine has been modified in such a way
that at least one charm quark appears in the final state in the same way as we
did for charm jets in Ref. \cite{3}.

The routine \cite{11, 12} is based on massless quarks, i.e. the charm quark is
also assumed massless. This is justified as the transverse momentum $p_T$ of the
produced charm jet is large enough, i.e. $p_T \gg m_c^2$, which is the case for
the two data sets of CMS in \cite{10}.

For our predictions of the inclusive c-jet cross-section we employ the MSTW
2008 NLO \cite{13} PDF (central value) of the Durham collaboration as we did in
our earlier work for bottom jets \cite{4}. The chosen asymptotic scale
parameter $\Lambda^{(5)}_{\overline{MS}} = 0.226$ GeV corresponds to 
$\alpha_s^{(5)}(m_Z) = 0.118$. The $\Lambda^{(5)}_{\overline{MS}}$ is adjusted to 
the appropriate $\Lambda^{4)}_{\overline{MS}}$ value for calculating charm. We 
shall also check for comparison two more modern PDF sets, namely CT14 \cite{14}
and MHSW \cite{15}. The center-of mass energy of the proton-proton collisions 
is taken $\sqrt{S} = 2.76$ and $\sqrt{S} = 5.02$ TeV as in the CMS data sets.
The comparison of results for the different PDFs is done only for 
$\sqrt{S} = 5.02$ TeV. We choose the renormalization scale 
$\mu_R = \xi_R p_T$ and the factorization scale $\mu_F = \xi_F p_T$ where $p_T$
is the largest transverse momentum of the two or three final state jets.
$\xi_R$ and $\xi_F$ are dimensionless scale factors, which are varied around 
$\xi_R =\xi_F$ to be specified later.

\section{Results and Comparison with CMS Data}

As a check of our program we have already calculated the cross-section
$d\sigma/dpp_Tdy$ for $pp \to single~jet + X$ for $\sqrt{S} = 7$ TeV in the 
$p_T$ region $18 \leq p_T \leq 1684$ GeV and in the rapidity region 
$|y| \leq 0.5$ and compared to CMS data \cite{16} in $p_T$ bins as chosen by 
CMS \cite{16}. Nonperturbative (NP) corrections for hadronization and multiple 
parton interactions were estimated by CMS \cite{16} and are published in the 
Durham Hep Data project \cite{17}. They are applied to the NLO perturbative QCD 
predictions. These corrected inclusive jet cross-sections including the 
theoretical error together with the CMS data are shown in Fig. 1 of our work 
\cite{3}. The agreement between data and the theoretical prediction shown there 
is quite good. The PDF used in this comparison was CT10 \cite{18}.
\begin{figure*}
\includegraphics[width=7.5cm]{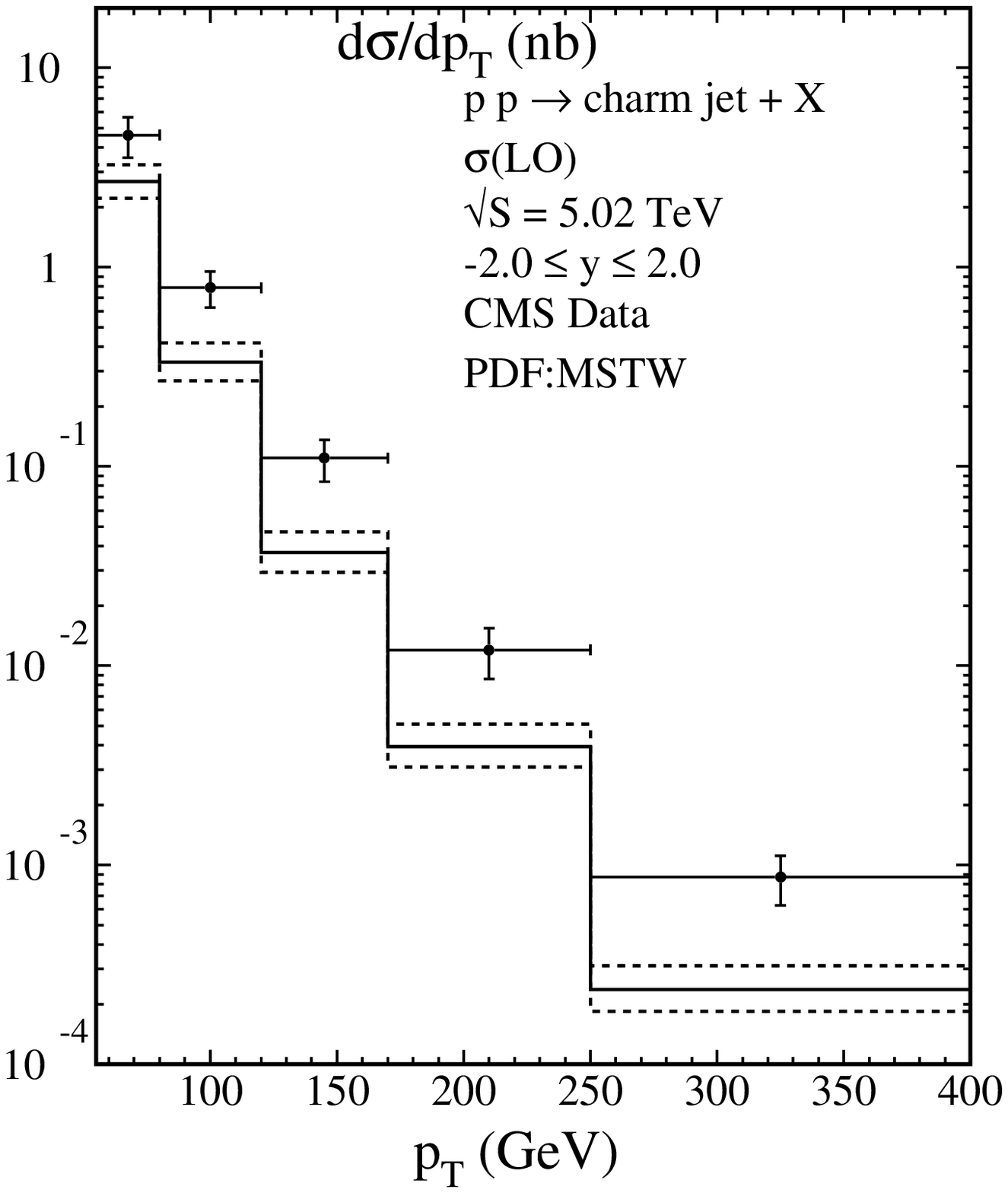}
\includegraphics[width=7.5cm]{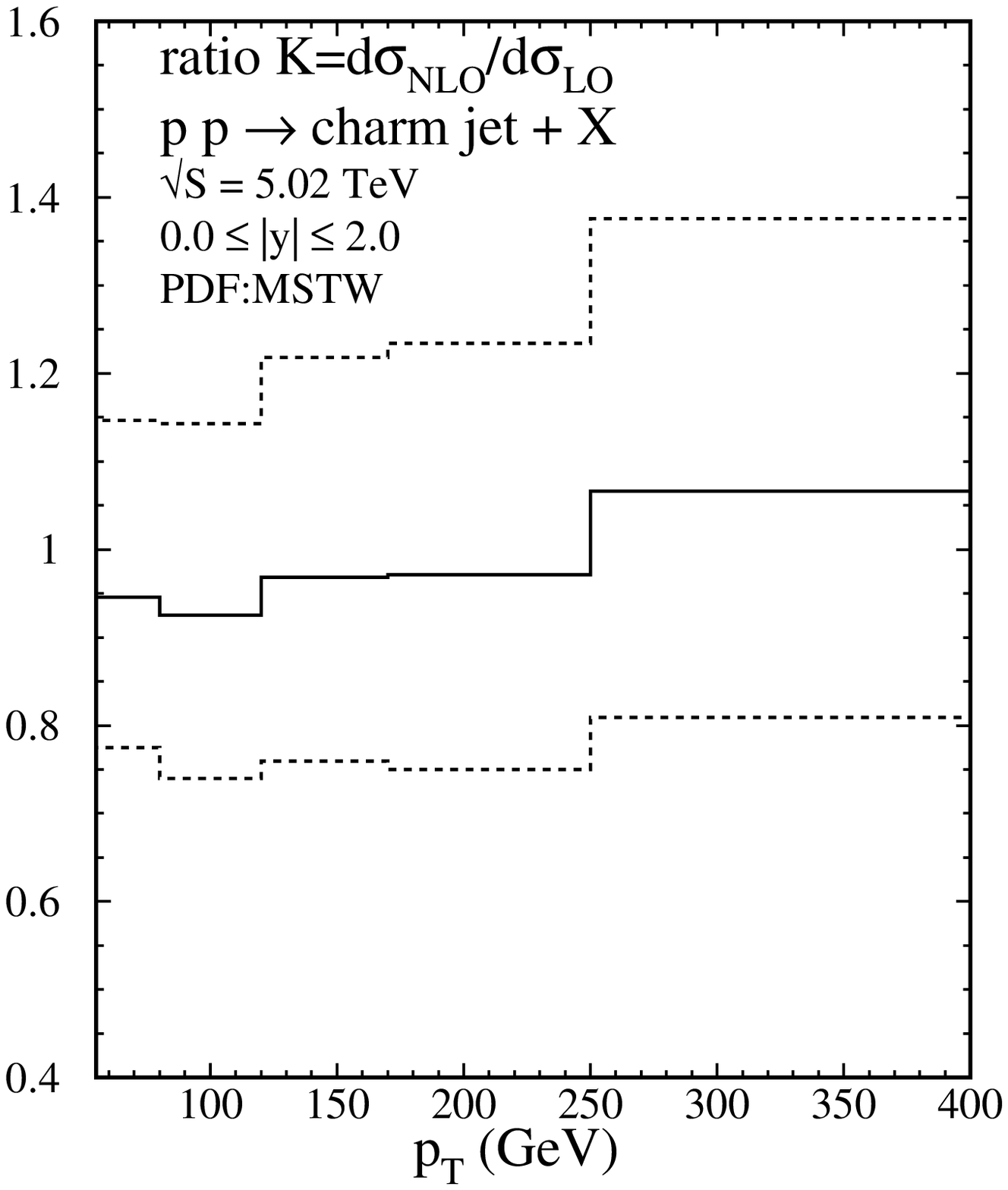}
\caption{\label{fig:1} Left side: single-inclusive c-jet cross section in
LO as a function of $p_T$ integrated over the rapidity region $0.0 \leq |y| \leq
2.0$ compared to CMS data \cite{10}, which are the points with error bars. 
The LO theoretical predictions are not corrected
by non-pertubative effects via a multiplicative factors. The theotetical error
(dashed lines) is obtained by scale variation as given in the text. The solid
line indicates the default scale choice. Right side: Ratio of single-inclusive 
c-jet cross section in NLO and LO integrated over rapity in the region
$0.0 \leq |y| \leq 2.0$ as a function of $p_T$ for three scale choices as given
in the text.}
\end{figure*}
Concerning the charm-jet calculation we have first calculated the inclusive 
cross-section $d\sigma/dp_T$ integrated over $|y| \leq 2.0$ for $\sqrt{S} =
5.02$ TeV in LO pQCD for the five $p_T$ bins as in the CMS paper \cite{10}. 
The scales for the LO cross-section are chosen as $\xi_R = \xi_F = 1.0$
(default solid line histogram in Fig.1, left side), $\xi_R = \xi_F = 0.50$
(upper dashed line histogram in Fig 1,left side) and $\xi_R = \xi_F = 2.0$
(lower dashed line histogram in Fig 1, left side) following our previous work 
\cite{4}. These three
calculations are compared to the inclusive charm-jet cross-section in Fig. 1,
left side for the same $\sqrt{S}$. These theoretical LO cross-sections are not 
corrected for NP effects. In addition, in the right frame of Fig. 1 we show 
the K factor, the ratio of the NLO to the LO cross-section, where 
$K = (d\sigma/dp_T)_{NLO}/(d\sigma/dp_T)_{LO}$ as a function of $p_T$ for the 
five $p_T$ bins as in the CMS data for $\sqrt{S} = 5.02$ TeV and for the same 
scale choices as in the left frame of Fig. 1. The NLO cross-section in the 
denominator is calculated with the scale $p_T$ and is used for the ratio with 
the three LO cross-section. As we shall see later the NLO cross-section depends
much less on the choice of scales. In the left frame these cross
section results are compared to the measured CMS values \cite{10}. As we can 
see, the calculated LO cross-sections are smaller than the data by 
approximately a factor of 1.8 for the lowest $p_T$ bin and a factor 3.7 for 
the largest $p_T$ bin. For jet cross-sections of all flavors, the NLO 
corrections usually increase the corresponding LO cross sections by factors 2 
to 3. If this would be the case also for thre c-jet cross-sections we would 
have approximate agreement with the data.
The corresponding K factor is shown in Fig. 1 (right frame). It is approximately
equal to $1.0 \pm 0.3$. The error is due to the scale dependence of the LO 
cross-section $d\sigma/dp_T$. It is nearly constant as a function of $p_T$.
This is in contrast to the b-jet production K factor in Ref. \cite{4}, which 
increased from 0.6 to 1.2 in the same $p_T$ range as for the c-jet 
cross-section $d\sigma/dp_T$. Results for $\sqrt{S} = 2.76$ TeV, not shown in 
this work, are expected to look similar.

Next, we want to show, how our NLO predictions describe the CMS c-jet cross-
section data \cite{10}. For this comparison we need the non-perturbative
corrections. For this we take the PYTHIA prediction also contained in Ref. 
\cite{10} for $\sqrt{S} = 2.76$ TeV and $\sqrt{S} = 5.02$ TeV. Unfortunately 
these predictions are not given in numerical form. So, we read them from Fig. 6
(upper and lower frame) in Ref. \cite{10}. Although errors for these 
cross-sections are shown in 
these figures also, it is difficult to obtain them from the figures. From 
the PYTHIA cross sections calculated by CMS from the PYTHIA routine we subtract
our LO predictions with the scale choice $\xi_R = \xi_F = 1.0$. The result is 
shown in Fig. 2 (left frame) for $\sqrt{S} = 2.76$ TeV and in Fig. 3 (left 
frame) for $\sqrt{S} = 5.02$ TeV as dotted histograms. In these two figures we 
show also the two CMS data sets for the two center-of mass energies and the 
corresponding NLO predictions for $d\sigma/d_T$ for the respective $p_T$ bins 
and in the rapidity range $|y| \leq 2.0$  as for the CMS data. These
predictions are shown for three scale choices (full histogram for default, upper
and lower dashed histograms for maximal and minimal cross sections). The
selection of these scales was performed as follows. First we calculated the NLO
cross sections for both $\sqrt{S}$. values by varying the scales according to 
the following combinations of $\xi_R$ and $\xi_F$: (1,1), (1,2), (2,1), (2,2), 
(1/2,1), (1,1/2) and (1/2,1/2). The largest up and down cross-sections for 
these choices are taken as the scale variation. The default cross-section is 
taken as the middle value of the maximal and the minimal cross-section out of 
these seven scale choices. It is found that the changes of these cross-sections 
between the maximal and minimal scale choices are rather small, and much 
smaller than in the LO predictions. This is understandable because of the 
well-known compensation of scale errors in the NLO case.
\begin{figure*}
\includegraphics[width=7.5cm]{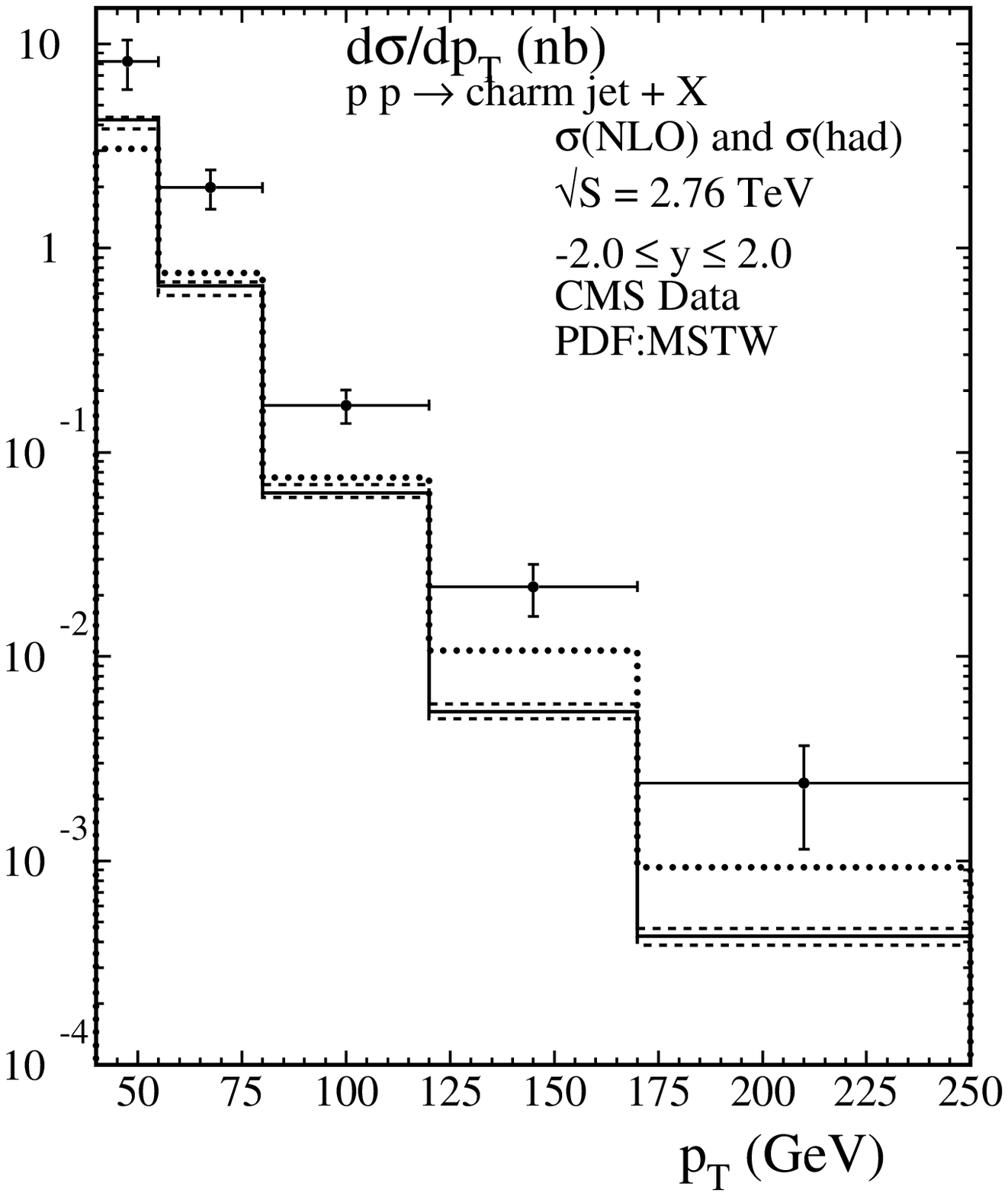}
\includegraphics[width=7.5cm]{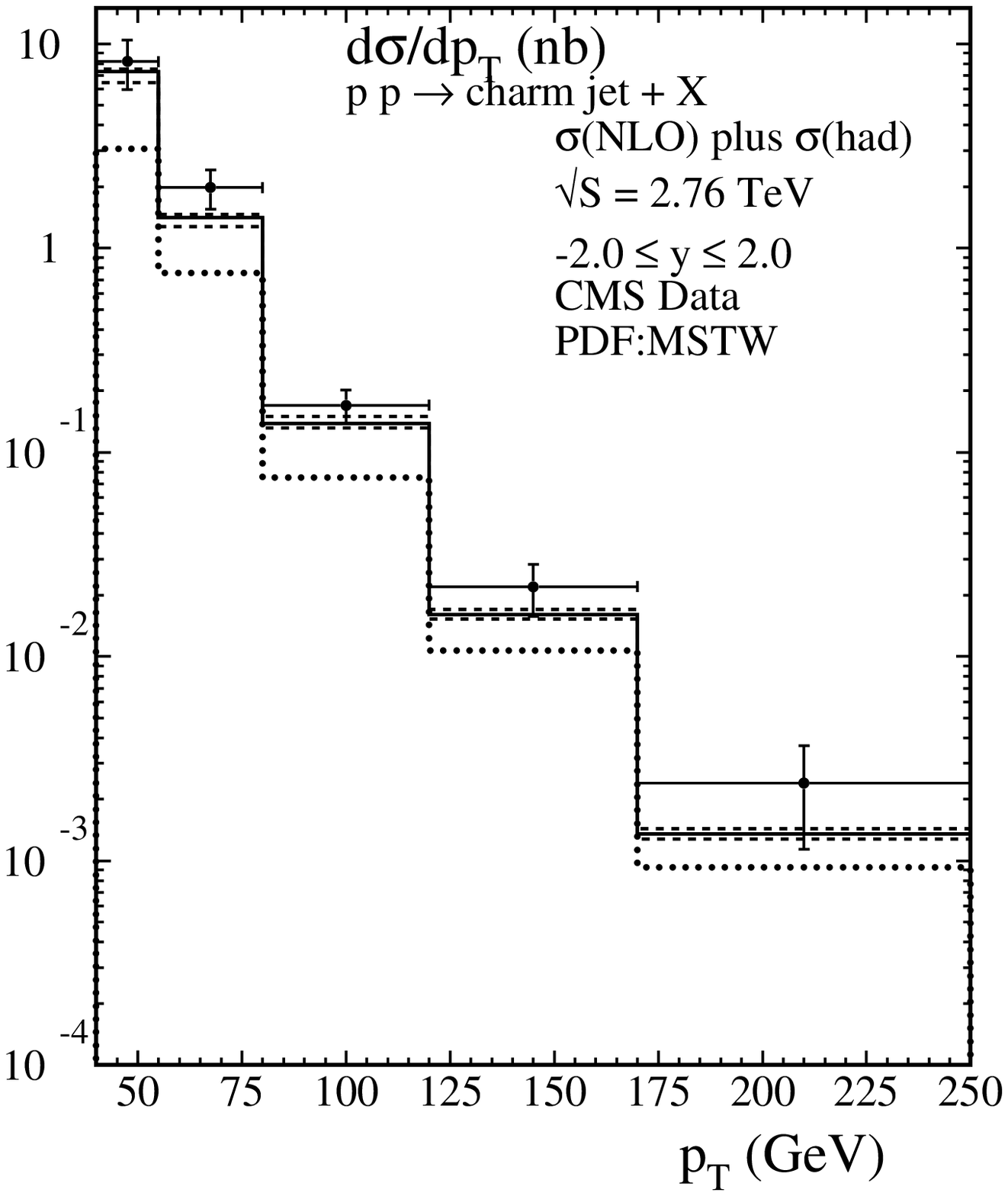}
\caption{\label{fig:2} Left side: Single-inclusive c-jet cross-section in NLO as
a function of $p_T$ integrated over the rapidity region $0.0 \leq |y| \leq
2.0$ at $\sqrt{S}= 2.76$ TeV compared to CMS data \cite{10}, given by the full 
line histogram and points with error bars. The NLO theoretical predictions is 
not corrected by non-pertubative effects via multiplicative factors. The 
theoretical error (dashed histograms) is obtained by scale variation as given 
in the text. The solid histogram indicates the default scale choice. The dotted
histogram is the PYTHIA minus LO prediction. Right side: Single-inclusive c-jet
cross-section in NLO plus the PYTHIA minus LO prediction as
a function of $p_T$ integrated over the rapidity region $0.0 \leq |y| \leq
2.0$ compared to CMS data \cite{10}, given by the full lines and points 
with error bars.}
\end{figure*}
\begin{figure*}
\includegraphics[width=7.5cm]{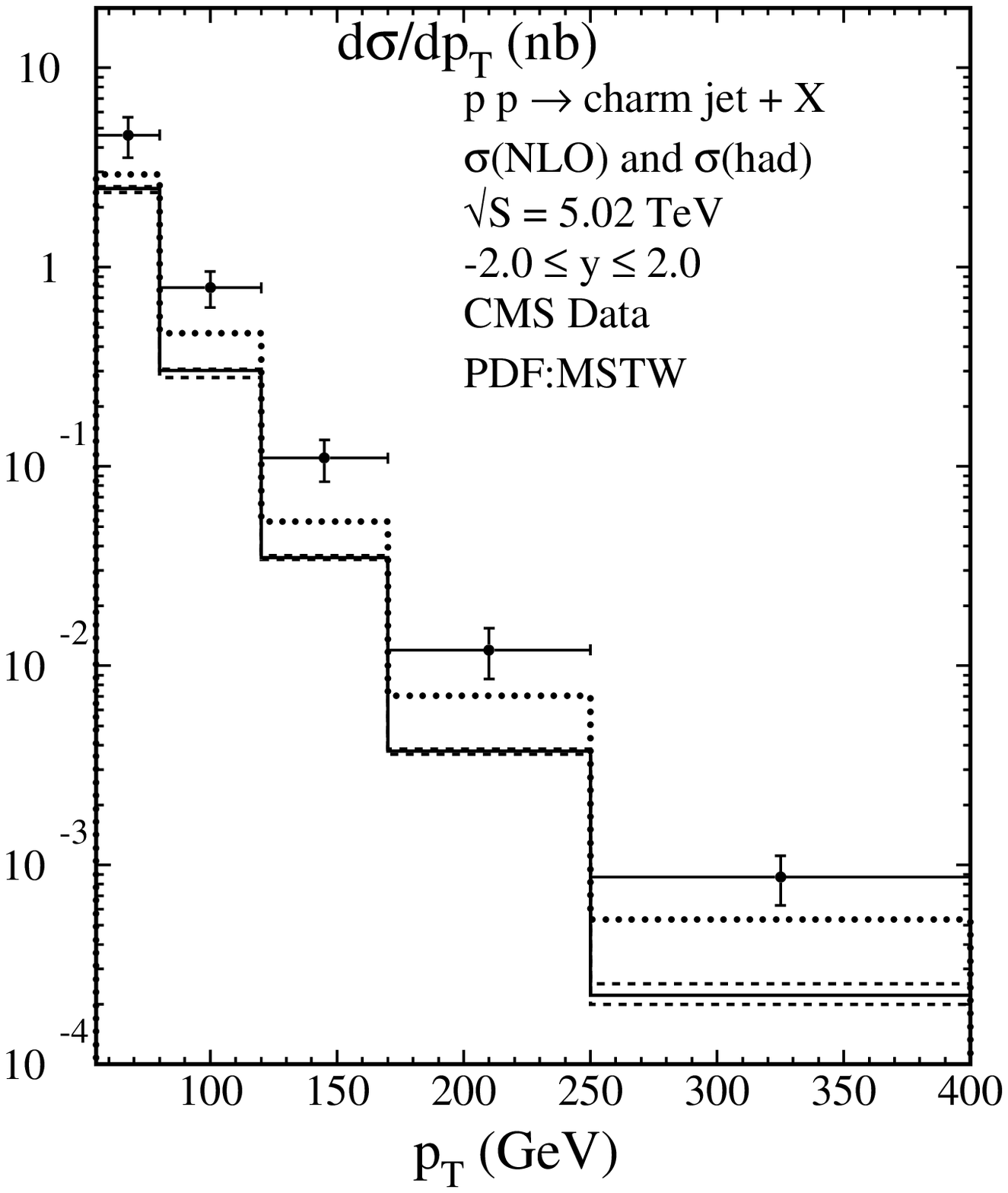}
\includegraphics[width=7.5cm]{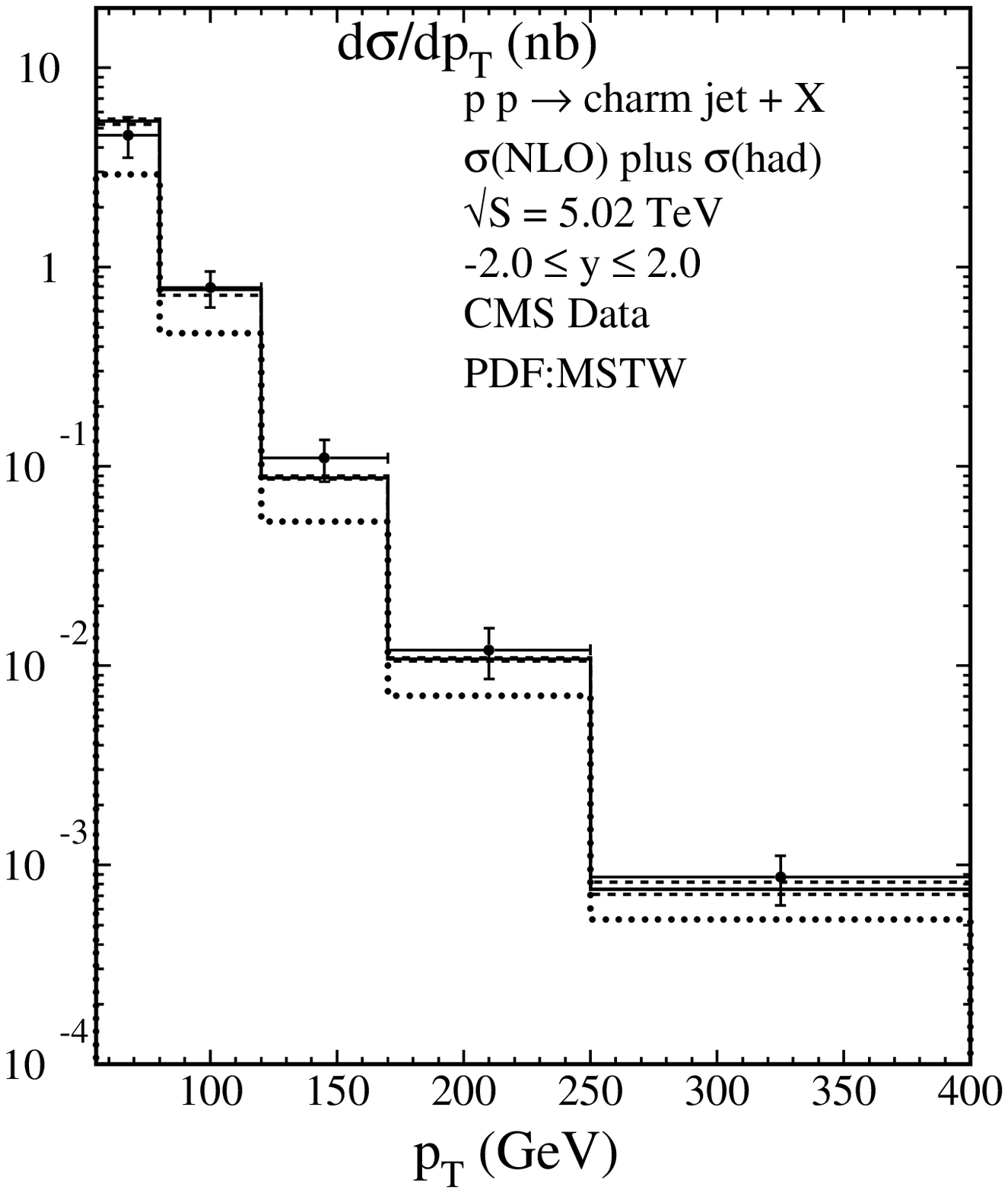}
\caption{\label{fig:3} Left side: Single-inclusive c-jet cross-section in NLO as
a function of $p_T$ integrated over the rapidity region $0.0 \leq |y| \leq
2.0$ at $\sqrt{S} = 5.02$ TeV compared to CMS data \cite{10}, given by the full 
line histogram and points with error bars. The NLO theoretical predictions is 
not corrected by non-pertubative effects via multiplicative factors. The 
theoretical error (dashed histograms) is obtained by scale variation as given 
in the text. The solid histogram indicates the default scale choice. The dotted
histogram is the PYTHIA minus LO prediction. Right side: Single-inclusive c-jet
cross-section in NLO plus the PYTHIA minus LO prediction as
a function of $p_T$ integrated over the rapidity region $0.0 \leq |y| \leq
2.0$ compared to CMS data \cite{10}, given by the solid lines and points 
with error bars.}
\end{figure*}
Our main results are shown in Fig. 2 $\sqrt{S} = 2.76$ TeV and Fig. 3 
$\sqrt{S} = 5.02$ TeV, respectively. In the left frame of these two figures we 
show our result for the NLO cross sections $d\sigma/dp_T$ for the five $p_T$ 
bins and $|y| \leq 2.0$ for the two $\sqrt{S}$ values, respectively, together
with the corresponding CMS \cite{10} data for these cross sections and the
“PYTHIA” cross section $d\sigma/dp_T$ read from Fig. 6 of Ref. \cite{10} minus 
the LO cross section $d\sigma/dp_T$ for the default scale with $\xi_R = \xi_F =
1.0$ as the  dotted histogram. We observe that the NLO cross-section values in 
Fig. 3. (left frame) differ only little from the corresponding LO cross section
value. This we should have expected from the K factor shown in Fig. 1 (right 
frame). In addition, we observe that the theoretical error due to the scale 
variations is very much reduced in the NLO predictions in Fig. 3 (left frame) 
when compared to the LO predictions in Fig. 1 (left frame). The next step is 
to add the PYTHIA cross-section minus the LO cross-section  to the NLO 
theoretical cross section as given by the dotted
histogram. The result can be seen in Fig. 2 and 3 (right frame), respectively.
As is seen, we obtain rather good agreement between our prediction including
non-perturbative correction and our NLO predictions for $d\sigma/d_T$. In 
Fig. 2 (right frame) at $\sqrt{S} = 2.76$ TeV for four $p_T$ bin cross-sections
$d\sigma/dp_T$  (out of five) we have agreement inside the experimental errors 
and in Fig. 3 (right frame) all five $p_T$ bin cross-sections $d\sigma/dp_T$ 
agree with the CMS data inside the experimental accuracy. This agreement is not
unexpected. In the CMS publication \cite{10} it was shown that the measured 
cross-section $d\sigma/dp_T$ for both center-of-mass energies agree quite well 
with the PYTHIA 6.424 \cite{19}, tune Z2 \cite{20} prediction. In our
results shown in Fig. 2 and Fig. 3 (right frame), respectively, the LO cross
section contained in PYTHIA 6 is just replaced by our NLO $d\sigma/dp_T$. Since
they are almost equal as shown in Fig. 1, (right frame) it is clear that our 
prediction of the NLO cross section plus the hadronic correction taken from 
PHYTHA must agree with the CMS data. In order to account for a possible scale 
dependence of the PYTHIA minus LO cross section we have in Fig. 2 and Fig. 3 
(right frame) doubled the theoretical error due to scale variation obtained in 
the NLO cross-section. As is seen in Fig. 2 (left frame) and Fig. 3 (left frame)
the correction from PYTHIA denoted $d\sigma(had)/dp_T$ (the dotted histogram) 
is either equal or larger than $d\sigma(NLO)/dp_T$, so that for most of the 
$p_T$-bins the PYTHIA correction dominates the experimental cross-section.
 
We have checked that our results do not depend on the choice of the NLO
PDF MSTW. For this purpose we have replaced the MSTW PDF by the NLO Ct14 
\cite{14} and NLO MMHT \cite{15} PDFs. We did not repeat all the calculations 
of cross sections as shown in Fig. 1, 2 and 3 for MSTW. We calculated only
the LO cross-sections as shown in Fif. 1 (left frame) and the K-factor in Fig.1
(right frame). For the latter we needed also the NLO cross-sections. These four
cros-sections were calculated with the same choice of scales as for the
MSTW PDF described above. The results for the other two PDFs differed only
by a few percent. Of course the LO cross-sections differed somewhat less than
the NLO cross ections needed for the K-factor. This K-factor for the CT14 and
MMHT PDFs showed the same behaviour as for the MSTW PDF in Fig 1 (right side).
In particula $K \simeq 1 $ for the five $p_T$ bins as in Fig 1 (right frame).
These calculations have been done only for the larger $\sqrt{S} = 5.02$ TeV. 
We expect very similar results for the lower $\sqrt{S}$ value. Of cource
we can expect that our results in Fig. 2 and 3 (right side) will look similar
for the CT14 and MMHT PDFs if the same choice of scales as for the MSTW PDF 
will be assumed.

A quite similar study has been done in our earlier work \cite{4} on single 
bottom jets and for the MSTW PDF. Also in this case we obtained
rather good agreement between CMS data at $\sqrt{s} = 7$ TeV and in two 
rapidity ranges $|y| \leq 0.5$ and $0.5 \leq |y| \leq 1.0$ after correcting the
calculated NLO cross-section $d\sigma/dp_T$ with $d\sigma(had)/dp_T$  as
calculated in the charm case form the PYTHIA prediction which gave a reasonable
description of the experimental CMS single-bottom jet cross section for 
$p_T \geq 50$ GeV shown in \cite{21}. Also in the bottom case the PYTHIA 
corrections yielded the largest part to the measured cross section. The K 
factor ($K = d\sigma(NLO)/d\sigma(LO)$ was also calculated in Ref.\cite{4}, 
but only for the CDF energy $\sqrt{S} = 1.96$ TeV for $p\bar{p} \to b-jet+X$ 
and for $|y| \leq 0.7$. It came out as $K \simeq 1.0$ but varying as a 
function of $p_T$ in the region $50 \leq p_T \leq 400$ GeV between 
$0.5 < K < 1.3$. We expect similar results for $pp$ collisions at$\sqrt{S} = 7$
and 13 TeV.

It seems that the hadronic corrections in the PYTHIA predictions account only
for a moderate correction to the NLO cross section, so that the major part must
originate from the parton-shower corrections. This has been confirmed in a more
recent analysis of inclusive b-jet production cross-sections measured by the CMS
collaboration at $\sqrt{S} = 13$ TeV described in the thesis by P. Connor 
\cite{22}. His work is based on CMS measurements of b jets in the $p_T$ range 
from $p_T = 74$ GeV to $1$ TeV for several rapidity bins in the range from 
$-2.4$ to $+2.4$. These data have much higher statistics than the earlier CMS 
and ATLAS b-jet measurements. He compared the b-jet cross-sections from the 
analysis of the more recent CMS data with two theoretical approaches, a LO 
routine with PYTHIA 8.1 corrections, which contains also parton-shower and 
parton-hadron corrections as the earlier PYTHIA Monte-Carlo programs. The 
second approach he compares the data to is a NLO routine with PYTHIA 8.1 
corrections, denoted POWHEG \cite{24}. Both routines give very
good agreement with the new CMS data. This shows in particular, that for the
single b-jet production in LO and NLO the parton-shower contributions are very
important in contrast to the single inclusive jet-production of light quark 
flavours where this is not the case. In Ref. \cite{22} it is demonstrated 
explicitly, that the parton-shower in PYTHIA dominates the b-jet production 
and the so-called hadronic corrections in PYTHIA are of minor importance. 
The reason for this difference between light-quark and b-jets is attributed to 
the fact that in b-jet production the LO process is already dominated by the 
gluon-splitting production into $b\bar{b}$. This is not the case in 
light-quark production, which is dominated by other LO processes.

We assume.that the parton shower corrections are also dominant in the case of
the charm-jet production similar to the bottom-jet production, although this was
not explicitly investigated in Ref. \cite{22}. For both cases, c-jet and b-jet,
the question arises whether the main part of the parton shower corrections 
comes from the NNLO contributions of the parton shower or whether even
higher parton showers or the whole sum is needed to obtain the whole PYTHIA 
corrections. In the latter case this would mean that the heavy-quark jet
cross-sections can not be calculated by pertubative fixed-order QCD.

It is clear, that our approach contains an inconsistency with respect to the NLO
corrections. On the one hand they are contained in the full NLO corrections
calculated by us and on the second hand they are contained in the PYTHIA
corections in the parton shower approximations. This problem has been
investigated by Connor in his thesis for the case of single inclusive b-jet
production \cite{22} by comparing the CMS data with the NLO POWHEG \cite{24}
routine showing as good agreement as for the pure PYTHIA comparison. We
expect a similar agreenment between these two approaches in the case of c-jet
production.

As a final point we mention that the ALICE collaboration has also measured
c-jet production in $pp$ collisions at $\sqrt{S} = 7$ TeV at the LHC \cite{25}.
The charm jets were identified by the presence of a $D^0$ meson
among the constituents of the jet with the  constraint $p_{T,D} > 3$ GeV. The
$D^0$-meson tagged jets are reconstructed using tracks of charged particles
only. Due to these additional constaints it is unclear how they can be
incorporated into the analytical calulations of the NLO calculation. Therefore
we did not attempt to compare our calculations with the ALICE data.

\section{Summary and Conclusions}

We have calculated the inclusive charm-jet cross section at NLO of QCD in the
ZM-VFN scheme, i. e. with active charm quarks in the proton at $\sqrt{S} = 2.76$
and $\sqrt{S} = 5.02$ TeV for $pp$ collisions at the LHC. The charm quarks
are considered massles. Our results are compared to experimental jet cross
section mesurements by the CMS collaboration at the LHC. To our surpise the
NLO cross sections are much smaller than the measured cross sections. They are
for the considered $p_T$ almost equal to the LO cross section, i. e. the K
factor is approximatel equal to one. There exist several possibilities to
increase this contribution. Examples are: contributions originating from
intrinsic charm-quark contributions to the proton PDF, as has been calculated
in \cite{3} or contributions fom higher orders than NLO from QCD, which may
become known in the future. This approach has been followed in this work by
approximating these higher order corrections by parton shower corrections
as contained in the PYTHIA Monte Carlo routines \cite{19}. It turns out, similar
to the results for b-jet production presented in \cite{4}, if the PYTHIA
predictions minus the LO pertubative cross section is added to the calcuted
NLO predictions reasonable agreement with the measured c-jet cross sections
can be achieved for CMS for both center of mass energies. In this connection the
question arises whether this mechanism might be applicable also for other
processses involcing heavy qarks, charm or bottom, as for example,
$pp \to c-jet + \gamma +X$ or $pp \to c-jet + Z +X$. It is well known that the
measurements of the first of these two examples is not in agreement with NLO
predictions \cite{26}.


\begin{thebibliography}{99}

\bibitem{1} S. Frixione, M.L. Mangano, P. Nason et al., Adv. Ser. Direct 
High Energy Phys. 15 (1998) 609; arXiv: hep-ph/9702287

\bibitem{2} S. Frixione and M.L. Mangano, Nucl. Phys. B483 (1997) 321
            arXiv: hep-ph/9605270

\bibitem{3} I. Bierenbaum and G. Kramer, Int. J. Mod. Phys. A30 (2015) 1550111
            arXiv:1412.5470 [hep-ph]


\bibitem{4} I. Bierenbaum and G. Kramer, Int. J. Mod. Phys. A31 (2016),
            1650098, arXiv:1603.01138 [hep-ph]

\bibitem{5} CDF and D0 Collaborations (M. D’Onofrio), arXiv:0505036 [hep-ex]

\bibitem{6} CDF and DO Collaborations (M. D’Onofrio) FERMILAB-CONF-06-224E

\bibitem{7} CDF Collaboration, CDF Note 8418, July 25, 2006

\bibitem{8} S. Chatrchyan et al., CMS Collaboration, J. High Energy Phys.1204
            (2012) 084, arXiv:1202.4617 [hep-ex]

\bibitem{9} G. Aad et al. ATLAS Collaboration, Eur. Phys. J C71 (2011) 1846,
            arXiv:1109.6833 [hep-ex]

\bibitem{10} A.M. Sirunyan et al., CMS Collaboration, Phys. Lett. B772 (2017),
             306, arXiv:1612.08972 [nucl-ex]

\bibitem{11} M. Klasen and G. Kramer, Z. Phys. C72 (1996) 107, 
             arXiv: hep-ph/9511405

\bibitem{12} M. Klasen and G. Kramer, Z. Phys. C76 (1997) 67, 
             arXiv: hep-ph/9611450

\bibitem{13} A.D. Martin, W.J.  Stirling, R.S. Thorne and G. Watt, Eur. Phys.
             J. C63 (2009) 189, arXiv: 0901,0002 [hep-ph]

\bibitem{14} S. Dulat et al., Phys. Rev D93 (2016) 033006, arXiv:1506.07443 
             [hep-ph]

\bibitem{15} L. A. Harland-Lang, A. D. Martin, P. Motylinski and R. S. Thorn,
             Eur. Phys. J C75 (2015) 204, arXiv:1412.3989 [hep-ph]

\bibitem{16} S. Chatrchyan et al., CMS Collaboration, Phys. Rev. Lett. 107
             (2011), 132001, arXiv: 1106,0208 [hep-ex]

\bibitem{17} http:hep data.cedar.ac.uk/view/ins 902309

\bibitem{18} H.L. Lai et al., Phys. Rev. D82 (2010) 074024, arXiv: 1007.2241
[hep-ex]

\bibitem{19} T. Sj\"{o}strand, S. Mrenna and P. Skands, J. High Energy Phys.
             05 (2006) 026, arXiv: hep-ph 0603175


\bibitem{20} R. Field, Acta Phys. Pol. B42 (2011) 2631, arXiv:1110.5530
             [hep-ph]

\bibitem{21} S. Chatrchyan et al., CMS Collaboration, J. High Energy Phys.,
 1204 (2012) 084, arXiv: 1202.4617 [hep-ex]

\bibitem{22} P. Connor, DESY-THESIS-2018-016

\bibitem{23} T. Sj\"{o}strand, S. Mrenna and P. Skands, Comp. Phys. Comm. 178
(2008) 852

\bibitem{24} S. Alioli et al., J. High Energy Phys. 1104 (2011) 081,
  arXiv:10122380 [hep-ph]

\bibitem{25} S. Acharya et al., ALICE Collabopration, J. High Energy Phys.,
  1808 (2019) 133, arXix: 1905.02510 [hep-ex]

\bibitem{26} V. M. Abazov,et al., D0 collaboration, Phys. Lett. B719
  (2013) 354, arXiv: 1210.5033 [hep-ex]

\end{thebibliography}
\end{document}